CAMBRIDGE
UNIVERSITY PRESS

**ARTICLE**

# Semi-empirical calibration of the oxygen abundance for LINER galaxies based on SDSS-IV MaNGA – The case for strong and weak AGN

C. B. Oliveira ●,[1] O. L. Dors ●,[1] I.A.Zinchenko ●,[2] M. V. Cardaci ●,[4] G. F. Hägele ●,[4] I. N. Morais ●,[1] P. C. Santos ●,[1] and G. C. Almeida ●[1]

[1]Universidade do Vale do Paraíba, Av. Shishima Hifumi, 2911, Cep 12244-000, São José dos Campos, SP, Brazil
[2]Faculty of Physics, Ludwig-Maximilians-Universität, Scheinerstr. 1, 81679 Munich, Germany
[3]Main Astronomical Observatory, National Academy of Sciences of Ukraine, 27 Akad. Zabolotnoho St 03680 Kyiv, Ukraine
[4]Facultad de Ciencias Astronómicas y Geofísicas, Universidad Nacional de La Plata, Paseo del Bosque s/n, 1900 La Plata, Argentina
[5]Instituto de Astrofísica de La Plata (CONICET-UNLP), La Plata, Avenida Centenario (Paseo del Bosque) S/N, B1900FWA, Argentina
**Author for correspondence:** C. B. Oliveira, Email: cbo_jr@hotmail.com.

**Abstract**

In this paper, we present a semi-empirical calibration between the oxygen abundance and the $N2$ emission-line ratio for Low Ionization Nuclear Emission Regions (LINERs). This relation was derived by comparing the optical spectroscopic data of 118 nuclear spaxels classified as LINERs using three different BPT diagrams from the Mapping Nearby Galaxies survey (MaNGA) and sub-classified as weak (wAGN, 84 objects) and strong (sAGN, 34 objects) AGN (active galactic nucleus) from the WHAN diagnostic diagram and photoionization model results obtained with the CLOUDY code assuming gas accretion into a black hole (representing an AGN). We found that our wAGN LINERs exhibit an oxygen abundance in the range of $8.50 \lesssim 12 + \log(O/H) \lesssim 8.90$, with an average value of $12 + \log(O/H) = 8.68$, while our sAGN LINERs exhibit an oxygen abundance in the range of $8.51 \lesssim 12 + \log(O/H) \lesssim 8.81$, with an average value of $12 + \log(O/H) = 8.65$. Our abundance estimations are in good agreement with those derived for another two different samples one of them with 463 Seyfert 2 objects and the other with 43 LINERs galaxies ionized by post-AGB stars, showing that the assumptions of our models are likely suitable for wAGN and sAGN LINERs. A relation between the equivalent width of the observed Hα emission-line and the estimated ionization parameter provided by models was obtained. Our results also suggest that LINERs does not show a clear correlation between oxygen abundances and the stellar mass of the hosting galaxies.

## 1. Introduction

Star-forming regions (SFs, i.e. H II regions and star-forming galaxies) and Active Galactic Nuclei (AGN) present strong emission lines in their spectra, which can be used to derive physical parameters of their ionized gas, such as electron densities ($N_e$), electron temperatures ($T_e$), metallicities ($Z$), etc (e.g., see Maiolino & Mannucci 2019 and Kewley et al. 2019 for reviews). Concerning the metallicity, generally, the oxygen abundance in relation to the hydrogen (O/H) is used as a tracer of it, since oxygen presents strong emission lines in the optical spectrum ([O II]$\lambda$3726,$\lambda$3729, [O III]$\lambda$5007) emitted by its most abundant ions (O$^+$, O$^{2+}$), according to studies of SFs (e.g. Kennicutt et al. 2003; Izotov et al. 2006; Dors et al. 2022) and AGN (e.g. Flury & Moran 2020; Dors et al. 2020a). Therefore, hereafter we use metallicity ($Z$) and oxygen abundance (O/H) interchangeably.

Basically, there are two different methods to derive the metallicity through emission lines. The first one is the $T_e$-method (for a review of the $T_e$-method see Peimbert et al. 2017 and Pérez-Montero 2017). Briefly, this method is based on direct determinations of electron temperatures, which requires measurements of auroral emission lines, such [O III]$\lambda$4363 and [N II]$\lambda$5755 (see Pilyugin 2003; Hägele et al. 2006; López-Sánchez & Esteban 2009; Hägele et al. 2008, 2011, 2012; Toribio San Cipriano et al. 2017; Hogarth et al. 2020). Unfor-tunately, auroral lines are weak (about 100 times weaker than Hβ) or even not detectable most objects with high metallicities and/or low excitation (van Zee et al. 1998; Díaz et al. 2007; Hägele et al. 2007, 2009, 2010, 2013; Dors et al. 2008).

Alternatively, a second method involving strong observational emission-lines has to be used when the $T_e$-method can not be applied. This kind of indirect method can be used to derive $Z$, as initially proposed by Pagel et al. (1979), who followed the original idea of Jensen et al. (1976). Indirect or strong-line methods are based on calibrations between strong emission-line ratios, easily measured in SFs and AGN spectra, and the metallicity (for a review on strong-line methods for SFs see López-Sánchez & Esteban 2010 and for AGN see Dors et al. 2020b). Over decades, several studies have proposed calibrations to derive metallicities in SFs (e.g. Pagel et al. 1979; Alloin et al. 1979; McGaugh 1991; Kewley & Dopita 2002; Pettini & Pagel 2004; Marino et al. 2013; Morales-Luis et al. 2014; Brown et al. 2016; Pilyugin & Grebel 2016; Curti et al. 2017; Ho 2019; Mingozzi et al. 2020; Pérez-Montero et al. 2021; Florido et al. 2022; Díaz & Zamora 2022), being few studies dedicated to AGN ( e.g. Castro et al. 2017; Pérez-Montero et al. 2019; Carvalho et al. 2020; Dors et al. 2021; Dors 2021; Carr et al. 2023) and to Low Ionization Nuclear Emission-line Regions (LINERs).

Regarding LINERs, despite this class of objects appearing in ∼ 1/3 of galaxies in the local universe (Netzer, 2013), the main ionization mechanism of the gas of these objects is still an open problem in astronomy, making difficult the determination of their metallicity and chemical abundance studies for this class of galaxies (Storchi-Bergmann et al., 1998). Actually, Annibali et al. (2010), by using optical spectra of a sample (65 objects) of LINERs located in early-type galaxies estimated the O/H abundance assuming ionization by hot main sequence stars (using the calibration by Kobulnicky et al. 1999) and by accretion of gas into a central black hole (applying the AGN calibration proposed by Storchi-Bergmann et al. 1998). These authors found that the AGN calibration produces higher values (∼ 0.05 dex in average) for the oxygen abundances than those derived through hot stars calibration.

Recently, there has been a renewed interest in the determination of metal abundances in LINERs. Krabbe et al. (2021) derived the O/H abundance in the UGC 4805 nucleus applying distinct methods: (i) comparison between observational data and results of photoionization models assuming gas accretion into a black hole (representing an AGN), (ii) photoionization models assuming post-Asymptotic Giant Branch (post-AGB) stars with different temperatures as ionizing sources and (iii) extrapolation of the disk radial abundance gradient to infer the nuclear abundance (see also Pilyugin et al. 2004; do Nascimento et al. 2022). These authors found that, depending on the method adopted, discrepancies until ∼ 0.4 dex in O/H estimates are derived (see Table 2 by Krabbe et al. 2021). This value is of the order of the discrepancies when using different strong-line methods for SFs (see López-Sánchez et al. 2012). Also, Pérez-Díaz et al. (2021) compiled optical spectroscopic data of 40 LINERs taken from the Palomar survey (Ho et al., 1995, 1997) and 25 LINERs from Pović et al. (2016), observed at the Calar Alto Observatory, totaling a sample of 65 LINERs (z ∼ 0.1), and applied the HII-CHI-MISTRY code (Pérez-Montero 2014, hereafter HCM code) to derive the O/H and N/O abundances. Pérez-Díaz et al. (2021) found a range of $8.01 \lesssim 12 + \log(\text{O/H}) \lesssim 8.86$ for their LINERs sample. Along this paper, we consider the solar oxygen abundance derived by Grevesse et al. (2010), which is $12 + \log(\text{O/H})_\odot = 8.69$.

For the first time, Oliveira et al. (2022) proposed two semi-empirical calibrations to estimate the oxygen abundances of LINERs. These authors selected a sample with 43 LINERs according to the [O III]λ5007/Hβ versus [N II]λ6584/Hα diagnostic diagram proposed by Baldwin, Phillips, & Terlevich (1981), known as classic BPT (hereafter [N II]-diagram). These nuclei were also classified as retired galaxies, i.e., the ionization of the gas is probably due to post-AGB stars (see the discussion by Cid Fernandes et al. 2011). Oliveira et al. (2022) built a grid of photoionization models using the CLOUDY code (Ferland et al., 2013) and considering post-AGB stars with three different effective temperatures (50, 100, and 190 kK) as the ionizing source. The results of these photoionization models were compared with the observational data of their sample. From this comparison, these authors were able to derive two semi-empirical calibrations, considering the $N2 = \log([\text{N II}]\lambda6584/\text{H}\alpha)$ and the $O3N2 = \log[([\text{O III}]\lambda5007/\text{H}\beta)/([\text{N II}]\lambda6584/\text{H}\alpha)]$ indexes as metallicity tracers. Through the proposed calibrations, these authors found that their LINERs present oxygen abundance values in the $8.48 \lesssim 12 + \log(\text{O/H}) \lesssim 8.84$ range, with an average value of $12 + \log(\text{O/H}) = 8.65$.

Recently, Oliveira et al. (2024) investigated the nitrogen abundances in the same LINER sample studied by Oliveira et al. (2022). These authors built detailed photoionization models with CLOUDY code (Ferland et al., 2017) to reproduce a set of observational emission line intensities ratios of the sample. By these models, the authors found nitrogen abundances in the range of $7.62 \lesssim 12 + \log(\text{N/H}) \lesssim 8.57$, with a mean value of $12 + \log(\text{N/H}) = 8.05 \pm 0.25$ and an oxygen abundance range between $8.05 \lesssim 12 + \log(\text{O/H}) \lesssim 9.03$, with a mean value of $12 + \log(\text{O/H}) = 8.74 \pm 0.27$. The LINERs analyzed by Oliveira et al. (2024) are located in the higher N/O region on the N/O versus O/H diagram, showing an unexpected negative trend between these two parameters. The authors investigated some explanations reported in the literature for these deviations, however, they did not find any evidence to support these mechanisms.

As a subsequent study, in the present work, we investigate the oxygen abundance of LINER galaxies whose ionization sources are probably AGN, since these objects are classified as LINERs in the BPT diagram (Baldwin et al. 1981) and in strong and weak AGN (sAGN and wAGN, respectively) by using the WHAN diagram (Cid Fernandes et al., 2011). Following a similar methodology as the one applied by Oliveira et al. (2022), and also using observational data taken from Mapping Nearby Galaxies at APO (MaNGA Bundy et al. 2015) survey, in the present paper, we propose a new semi-empirical metallicity calibration for these objects by using the $N2$ index as a metallicity tracer. These object classes (sAGN and wAGN LINERs) were not considered by Oliveira et al. (2022) and could represent a substantial increase in the number of LINERs with metallicity determinations. Indeed, taking into account the available observational LINERs data in the MaNGA data set, the metallicity calibration for LINERs classified as wAGN and sAGN in the WHAN diagram and developed in the present study, makes it possible $Z$ estimates for a sample of 118 objects, increasing by a factor of about two the number of $Z$ estimates in LINERs in comparison to previous studies. In addition, metallicity discrepancies in LINERs ionized by distinct sources could reveal the effects of some physical processes on the enrichment of the interstellar medium (ISM), as the influence of gas outflows present in some LINERs classified as wAGN and sAGN (e.g. Ilha et al. 2022). We carried out our analysis using the CLOUDY code (Ferland et al. 2013) to build grids of photoionization models, representing AGN, to reproduce strong optical emission line ratios found for the objects in our sample.

This paper is organized as follows: Section 2 describes the observational data, photoionization models, and the methodology applied to derive the oxygen abundances. In Section 3, a comparison between the observational data and photoionization models as well as the calibration obtained are presented, while in Section 4 the discussion of the results is shown. In Section 5, we present our conclusions. Throughout this paper we adopt the Planck Collaboration et al. (2021) cosmologic parameters: $H_0 = 67.4$ km s$^{-1}$ Mpc$^{-1}$ and $\Omega_m = 0.315$.

## 2. Methodology

### 2.1 Sample selection

In this paper, we utilized optical spectroscopic data of LINERs obtained from the MaNGA SDSS DR17 survey (Abdurro'uf et al., 2022). To derive the emission line and continuum fluxes of each galaxy, we followed the methodology described by Zinchenko et al. (2019b, 2021) and Zinchenko (2023). Briefly, on each spectrum of our sample, the stellar component was fitted using the STARLIGHT code (see Cid Fernandes et al. 2005; Mateus et al. 2006; Asari et al. 2007), assuming as templates the simple stellar populations (SSPs) from the work by Bruzual & Charlot (2003). To fit the emission lines we used our ELF3D code, which was built upon the LMFIT package (Newville et al., 2016). Following Izotov et al. (1994), for each spectrum, we applied the Whitford reddening law analytical approximation (Whitford, 1958), assuming a Balmer line ratio of H$\alpha$/H$\beta$ = 2.86, which was obtained for recombination case B for an electron temperature of 10 000 K and an electronic density of 100 cm$^{-3}$. In spaxels with values of H$\alpha$/H$\beta$ lower than 2.86, we set the reddening to zero.

After we carried out this procedure, we obtained the reddening corrected spectra for the galactic nuclear spaxels, assuming a circular aperture with a diameter of ∼ 2 arcsec. For each object, we selected the nuclear spaxel with the highest signal-to-noise (S/N) of H$\alpha$. The signal-to-noise ratio (S/N) was required to be higher than 5 in all [O II]λ3727, H$\beta$, [O III]λ5007, [O I]λ6300, H$\alpha$, [N II]λ6584, and [S II]λ6716, λ6731 emission lines.

Subsequently, the BPT diagrams [O III]λ5007/H$\beta$ vs. [N II]λ6584/H$\alpha$, [S II](λ6716+λ6731)/H$\alpha$ and [O I]λ6300/H$\alpha$ (Baldwin et al. 1981 and Veilleux & Osterbrock 1987), to classify each nuclear spaxel in our sample were used. Initially, the empirical and theoretical criteria proposed by Kauffmann et al. (2003) and Kewley et al. (2006), respectively, to classify objects in H II-like regions, composite, and AGN-like objects were considered. Also, the criteria suggested by Cid Fernandes et al. (2010) to separate LINERs from Seyfert nuclei in BPT diagrams were used. However, it is difficult to discriminate the ionization source of LINERs only through the BPT diagrams (e.g. Stasińska et al. 2015). Thus, the WHAN diagram proposed by Cid Fernandes et al. (2010, 2011), which uses the equivalent width of H$\alpha$ (EW$_{H\alpha}$) versus [N II]λ6584/H$\alpha$, is a useful tool to distinguish the nature of the ionization source of objects according to the following classification criteria:

1. Pure star-forming galaxies: log([N II]/ H$\alpha$) < −0.4 and EW$_{H\alpha}$ > 3 Å. These objects have as ionization sources O and/or B stars.
2. Strong AGN: log([N II]/H$\alpha$) > −0.4 and EW$_{H\alpha}$ > 6 Å.
3. Weak AGN: log([N II]/H$\alpha$) > −0.4 and EW$_{H\alpha}$ between 3 and 6 Å.
   Such as the strong AGN, weak AGN have ionizing source radiation coming from accretion of gas into a black hole.
4. Retired Galaxies (RGs; i.e., fake AGN): EW$_{H\alpha}$ < 3 Å. These objects, probably, have as ionization source post-AGB stars.
5. Passive galaxies (PGs): EW$_{H\alpha}$ and EW$_{[N II]}$ < 0.5 Å. PGs are defined as those with very weak or undetected emission lines.

To exclude any ambiguous object, we only selected galaxies with simultaneously LINER classification in the three BPT diagrams and also AGN (strong or weak) classification in the WHAN diagram. The final sample is composed by 118 LINER galaxies: 84 wAGN and 34 sAGN, as classified by the WHAN diagram, with a wide range of stellar masses [9.0 $\lesssim \log(M_*/M_\odot) \lesssim$ 11.2] and redshifts 0.02 $\lesssim z \lesssim$ 0.12 (masses and redshifts were taken from the MaNGA survey). For sample, in Figure 1 (right top panel), the observed (in black) and synthetic (in red) spectra of the selected nuclear spaxel of the sAGN MaNGA 7990–12704 object belonging to our sample are shown. In the right bottom panel the pure emission spectrum, i.e. after the SSP subtraction, as well as some emission line identifications, are shown. In the left panel of Fig. 1, a composite image of this object with the IFU field overlapped is shown. In Fig. 2, the BPT and WHAN diagrams containing our final sample are shown, where strong and weak AGNs are discriminated by distinct colors, as indicated.

In Appendix 1, the reddening corrected emission line intensities (in relation to H$\beta$ = 1), the logarithm of EW$_{H\alpha}$, the reddening coefficient [$c$(H$\beta$)] and the absolute flux of H$\beta$ of for each nuclear spaxel in our sample are listed.

### 2.2 Photoionization models

We built a grid of photoionization models with version 17.02 of the CLOUDY code (Ferland et al., 2017), assuming a wide range of nebular parameters. These dust-free models are similar to those built by Carvalho et al. (2020) and Oliveira et al. (2022). In our models, a plane-parallel geometry is adopted, and the outer radius is assumed to be the one where the gas temperature falls to 4 000 K (default outer radius value in the CLOUDY code), with a constant electronic density along the radius. A brief description of the input parameters is presented in what follows.

1. Spectral Energy Distributions (SEDs): Oliveira et al. (2022) considered a sample of LINERs classified in the WHAN diagram as retired galaxies. The photoionization models built by these authors assumed post-AGB star atmosphere models by Rauch (2003) as SEDs. In the present work, the LINERs of the sample are classified as strong or weak AGN, i.e. the ionization is produced by radiation from gas accretion into a black hole with distinct rates. Thus, to represent AGN SEDs, a multi-component continuum, similar to that observed in typical AGN, was assumed. In our models, we assumed three different values for the slope of the SED as defined by Avni et al. (1980): $\alpha_{ox} = -0.8, -1.1$, and $-1.4$. Carvalho et al. (2020) showed that photoionization models assuming similar $\alpha_{ox}$ range are able to reproduce optical line ratios of a large sample of Seyfert 2 nuclei (see also

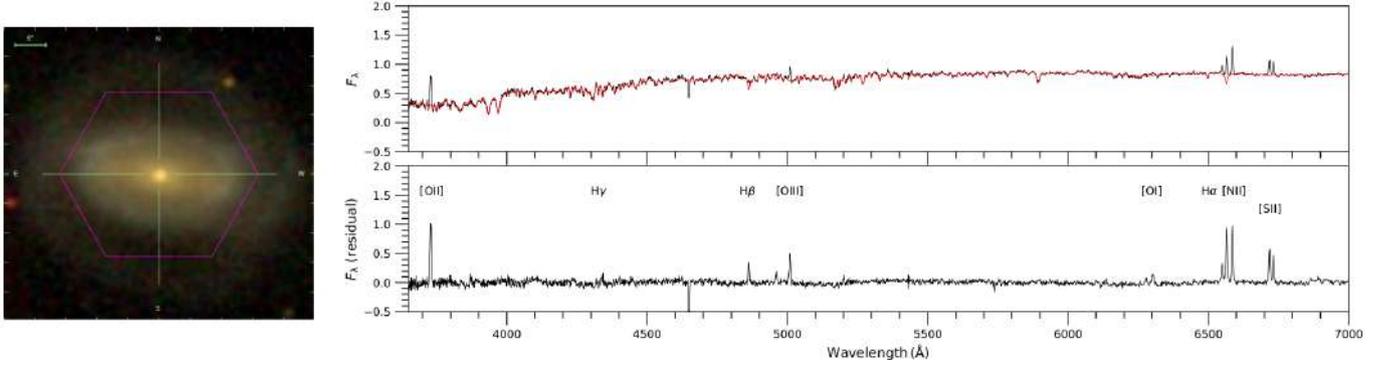

**Figure 1.** Left panel: SDSS *gri* band composite image of the nuclear spaxel of sAGN MaNGA 7990-12704 object taken from the MaNGA survey (Blanton et al., 2017). The IFU field of view is indicated in purple. Right upper panel: observed spectrum (in black) and its single stellar population synthesis (in red) for the selected spaxel of the MaNGA 7990-12704 object. Right lower panel: pure emission spectrum for the same object. with some emission lines identified

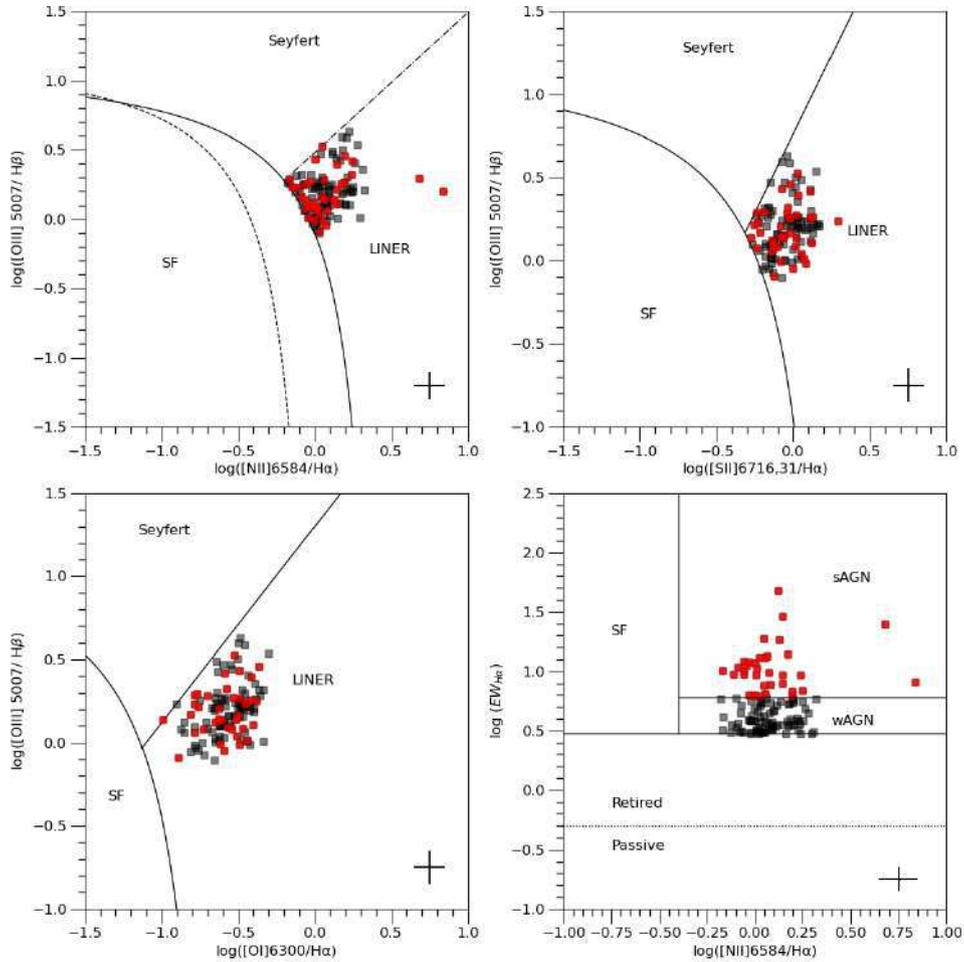

**Figure 2.** Top left panel: [O ɪɪɪ]λ5007/Hβ versus [N ɪɪ]λ6584/Hα diagnostic diagram. The black solid curve represents the theoretical upper limit for the star-forming regions proposed by Kewley et al. (2001), the black dashed curve is the empirical star-forming limit proposed by Kauffmann et al. (2003), and the pointed–dashed black line is the criteria proposed by Cid Fernandes et al. (2010) to separate LINERs from AGN. Top right panel: [O ɪɪɪ]λ5007/Hβ versus [S ɪɪ]λλ6716,31/Hα diagnostic diagram, with the criteria proposed by Kewley et al. (2006) to distinguish the objects. Bottom left panel: [O ɪɪɪ]λ5007/Hβ versus [O ɪ]λ6300/Hα diagram, with the criteria proposed by Kewley et al. (2006) to distinguish the objects. Bottom right panel: WHAN diagram. Black and red points represent the observational line ratios for the wAGN and sAGN nuclei, respectively, for the objects in our sample as classified by the WHAN diagram.

Dors et al. 2017a; Pérez-Montero et al. 2019). By using observational data, the value of $\alpha_{ox} \sim -1.0$ was derived as representative for LINERs and low luminosity AGN (see Ho 1999; Eracleous et al. 2010; Maoz 2007; Younes et al. 2012).

2. Metallicity: we assumed $(Z/Z_\odot)$ = 0.2, 0.5, 0.75, 1.0, 2.0, and 3.0 for the gas phase of the models. We also assumed the parametrization of Grevesse et al. (2010), in which the solar oxygen abundance $12 + \log (O/H)_\odot$ = 8.69 is equivalent to $(Z/Z_\odot)$=1.0. For nitrogen, we assumed the relation proposed by Carvalho et al. (2020), who considered abundance estimations for type 2 AGN and SFs. The other elements (e.g. S, Ar) were linearly scaled with the metallicity.

3. Electron Density: Oliveira et al. (2024) derived, by using photoionization models combined with observational data, a range of electron density ($N_e$) between $50 \lesssim N_e \lesssim 2800$ cm$^{-3}$, with an average value about $N_e \approx 400$ cm$^{-3}$. In the present work,we assumed electron density values $N_e$ = 100, 500, and 3 000 cm$^{-3}$, constant along the nebular radius. This is the same interval used by Oliveira et al. (2022) for photoionization model grids to reproduce observational data of LINERs classified as retired galaxies and used by Carvalho et al. (2020) to build a grid of models to reproduce data from Seyfert 2 nuclei.

4. Ionization Parameter: we followed Oliveira et al. (2022) and assumed the logarithm of $U$ in the range of $-4.0 \leq \log U \leq -1.0$, with a step of 0.5 dex. This is about the same range of values assumed by Feltre et al. (2016, 2023), who built a photoionization model grid to reproduce ultraviolet and optical emission-line ratios of AGN and SFs.

To estimate the metallicity and ionization parameters for the galaxies in our sample, we compared observational emission line intensity ratios with those predicted by our photoionization models. This comparison is done by using the diagrams: log([O III]λ5007/[O II] λ3727) versus $N2$, being [O III]/[O II] mainly sensitive to $U$ and $N2$ to $Z$.

Alternatively, the metallicity $Z$ can be estimated from bayesian-like comparison between certain observed and model-predicted emission-line ratios sensitive to total oxygen abundance (see Pérez-Montero 2014). For instance, Thomas et al. (2018) proposed a Bayesian code, called as NEBULABAYES, in which a comparison between observational data and photoionization model grids representing AGN is performed to estimate $Z$. This code determines the probability of a set of model parameter values (including a wide range of nebular parameters and SED of ionizing source) reproduces a given observational data. Thomas et al. (2019), applying the NEBULABAYES code, found that this analysis method produces similar (almost identical) results to those derived by Castro et al. (2017). The methodology proposed by Castro et al. (2017) is the same as that one applied by Dors et al. (2019); Carvalho et al. (2020); Oliveira et al. (2022), and, finally, by us in the present work. Thus, it is reasonable to assume that our grid of models seems to produce similar metallicity values to those derived from bayesian-like comparisons.

## 3. Results

Auroral emission lines (such as [O III]λ4363 and [N II]λ5755) are not measured in the spectra of the objects belonging to our sample, thus, it is not possible to apply the $T_e$-method, and $Z$ estimates are only possible by indirect or strong-line methods.

To obtain a strong-line method to estimate $Z$ in LINERs classified as sAGN and wAGN, we compare the results of AGN photoionization models to the observational data for the objects of our sample in the log([O III]λ5007/[O II] λ3727 versus $N2$ diagram (Fig. 3). This diagram combines the [O III]/[O II] lines ratio, mainly sensitive to the ionization degree of the gas (e.g. McGaugh 1991; Dopita et al. 2000), with $N2$, mainly sensitive to $Z$ (e.g. Storchi-Bergmann et al. 1994; Raimann et al. 2000). A similar methodology was recently applied by Carr et al. (2023), who compared observational data of Seyfert 2 nuclei and photoionization model results in a [O III]λ5007/Hβ versus $N2$ diagram. Dors et al. (2011) analyzed the reliability of oxygen abundance and ionization parameter estimates through a number of diagnostic diagrams containing SF photoionization results. The accuracy of the diagrams was analyzed by comparing O/H estimates from these with those via the $T_e$-method. Unfortunately, the $T_e$-method was not developed for LINERs and, thus, there are no (reliable) direct abundance estimates for this object class in the literature. Therefore, it is worth emphasizing that future direct O/H estimates are necessary to validate the abundance results obtained in the present work.

The log([O III]λ5007/[O II] λ3727) versus $N2$ diagrams containing the observational data and photoionization model results are shown in Fig. 3, where it can be seen that models with $\alpha_{ox} < -1.4$ well reproduce the observational data. It is also possible to note that models with $\alpha_{ox} = -1.4$ do not reproduce most parts of the observational data, in the sense that the $N2$ values predicted by the models are underestimated (by $\sim 0.5$ dex) in comparison to the observational ones (see also Dors et al. 2017a; Carvalho et al. 2020). Therefore, models with $\alpha_{ox} = -1.4$ are not considered in the derivation of the $Z-N2$ calibration.

To calibrate the $N2$ index as a function of the metallicity, we derived the logarithm of the ionization parameter and the metallicity for each galactic nucleus of our sample through linear interpolation between the models (see Dors et al. 2011). From this interpolation, for each object and for each diagram (i.e. for models with distinct $\alpha_{ox}$ and $N_e$ values, see Fig. 3), it is possible to obtain pairs of values ($Z$, $U$) and their corresponding ([O III]/[O II], $N2$) values. In this way, we derived a set of points for the sample, i.e. ($Z$, $U$)-([O III]/[O II], $N2$) and thus, a unidimensional $Z-N2$ calibration. In Fig. 4, the average values of the ($N2$, $Z$) obtained considering the six diagrams shown in Fig. 3 (values from models assuming $\alpha_{ox} = -1.4$ are not considered) are shown. A clear correlation between these quantities is found, with a correlation coefficient of $r = 0.97$, being represented by

$$12 + \log(O/H) = 0.72(\pm 0.09) x^2 + 0.68(\pm 0.01) x + 8.60(\pm 0.01). \tag{1}$$

being $x = N2$.

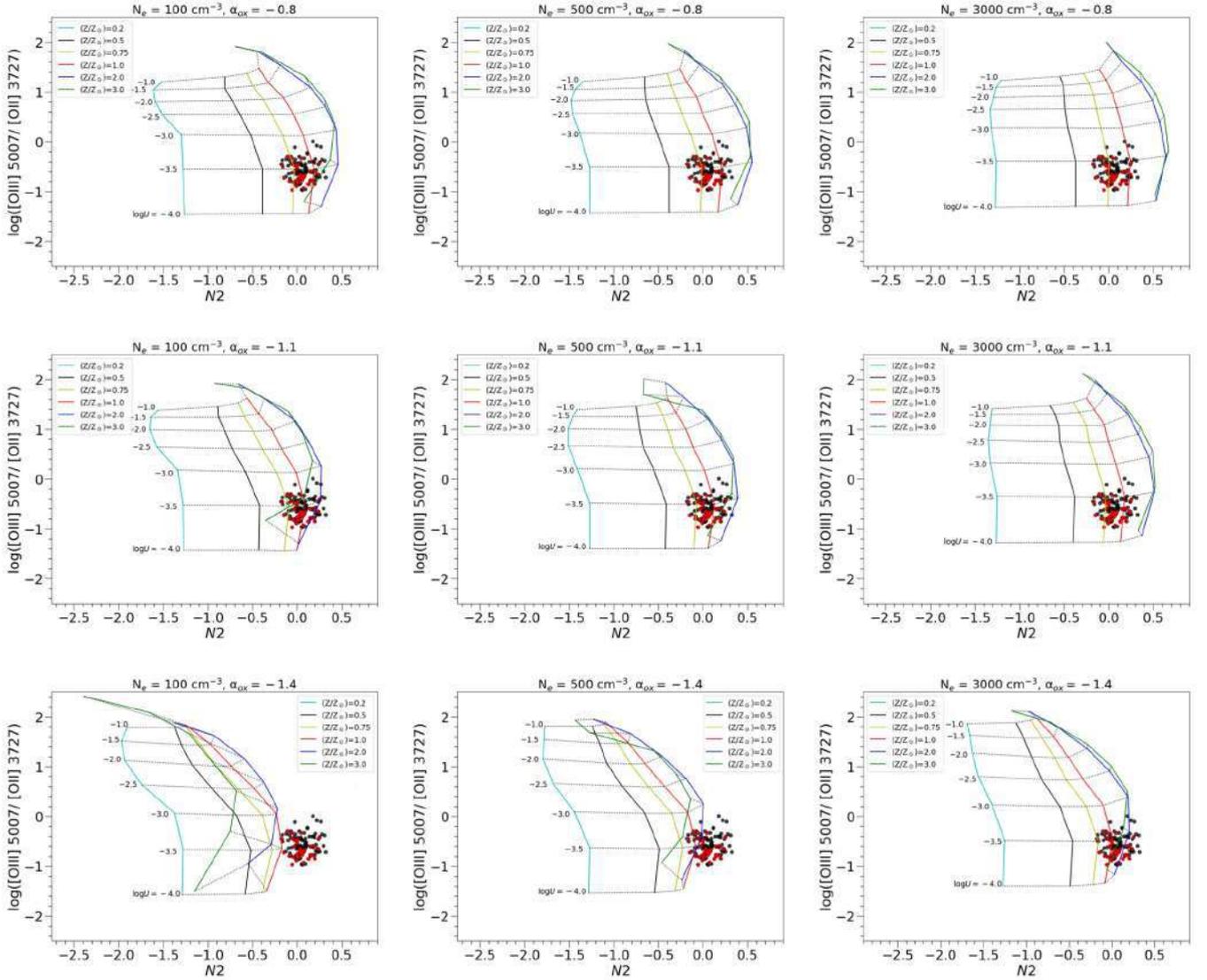

**Figure 3.** log([O III]λ5007/[O II] λ3727) versus *N*2=log([N II]λ6584/Hα) diagnostic diagram. Distinct colored solid lines connect the photoionization model (see Sect. 2.2) results with the same metallicity, while dotted lines connect models with the same ionization parameter (*U*), as indicated. Black and red points represent the observational line ratios (see Sect. 2.1) for each nucleus (wAGN and sAGN, respectively) of our sample. In each plot, a grid of models assuming different electron density (*N*ₑ) and αₒₓ values, as indicated, is shown.

In Fig. 4 this equation is represented by a red curve and it is valid for the range of $-0.2 < N2 < 0.35$.

From the interpolation between observational data and the results of photoionization model grids we derive oxygen abundance for the wAGN in the range of $8.50 \lesssim 12 + \log(O/H) \lesssim 8.90$ (i.e. $0.5 \lesssim Z/Z_\odot \lesssim 1.7$), with an average value of $< 12 + \log(O/H) >= 8.68 \pm 0.09$ (i.e. $< Z/Z_\odot >\sim 0.9$) while for the sAGN these quantities are $8.51 \lesssim 12 + \log(O/H) \lesssim 8.81$ (i.e. $0.5 \lesssim Z/Z_\odot \lesssim 1.31$) and $< 12 + \log(O/H) >= 8.65 \pm 0.07$ (i.e. $< Z/Z_\odot >\sim 0.8$). We assumed the solar oxygen abundance $\log(O/H)_\odot = -3.31$ derived by Grevesse et al. (2010). About $\sim 30$ per cent of the objects of our sample present oversolar abundances. It is worth mentioning that, as would be expected due to the $N2$ index mainly depending on metal abundances, both wAGN and sAGN are distributed along the entire metallicity range. See also how the distributions of both kinds of objects are overlapped in Fig. 3.

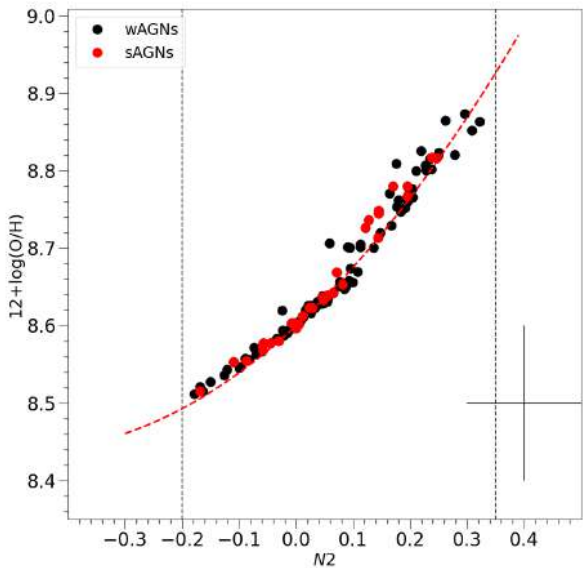

**Figure 4.** Oxygen abundance versus $N2$ parameter. Black and red points represent the average values of O/H (for wAGN and sAGN, respectively) derived from interpolation between observational data and the results of photoionization model grids from Fig. 3. The red curve represents the fitting to the points and it is given in Eq. 1, while the vertical dashed lines represent the valid interval for our calibration.

## 4. Discussion

To determine the metallicity of the gas phase of any object through its emission lines it is essential to know the nature of the ionizing source of the gas, especially when indirect methods are applied. Strong-line methods have been proposed over the years (see Maiolino & Mannucci 2019 and the references therein) to estimate $Z$ in SFs and AGN (mainly Seyfert 2s), however, an opposite situation is found regarding LINERs. The nature of the ionizing source of this class of galaxies is an open problem in astronomy and several works have pro-posed explanations for it. Heckman (1980) suggested shocks as responsible for the ionization of LINERs. On the other hand, radiation from gas accretion into black holes (AGN) was suggested as the ionizing mechanism of the gas (Ferland & Netzer 1983; Halpern & Steiner 1983; Ho et al. 1993). Also, hot evolved stars (like post-AGBs, e.g. Terlevich & Melnick 1985) or even normal main-sequence stars (O5 or earlier, e.g. Shields 1992) were proposed as the main ionizing source of LINERs. In the present work, we used three BPT diagnostic diagrams (Baldwin et al. 1981; Veilleux & Osterbrock 1987), as well as the WHAN diagram (Cid Fernandes et al., 2011), to select LINERs that are probably ionized by AGN. Therefore, based on this assumption, we proposed a semi-empirical calibration between the $N2$ line ratio and the metallicity derived from a comparison between photoionization models assuming an AGN as the main ionizing source of the gas and optical observational data (see Sect. 2).

### 4.1 Oxygen abundance: comparison with post-AGB LINERs and Seyferts nuclei

The results obtained in the present work, together with those from Oliveira et al. (2022), are the first metallicity estimates for LINERs taking into account the distinct ionizing source of this object class. Thus, a comparison between these estimates, as well as with those for confirmed AGN (see Dors et al. 2020b and references therein), produces important insights on metal ISM enrichment by stars within different environments. In fact, it is expected that in LINERs ionized by post-AGB stars the ISM enrichment would be, mainly, due to the metal releasing in the ISM by already evolved stars, i.e. there is not (or low level of) ongoing star formation. Otherwise, it has been observed some level of recent nuclear star formation in wAGN and sAGN LINERs and in "normal" AGN (e.g. Seyfert nuclei) (e.g. Shlosman et al. 1990; Storchi-Bergmann et al. 1996; Riffel et al. 2009, 2023). In addition, feedback processes in AGN ( e.g. see Fabian 2012 for a review.) and LINERs (e.g. Ilha et al. 2022) could suppress (e.g. Page et al. 2012; Barger et al. 2015), increase (e.g. Lutz et al. 2010; Rosario et al. 2012; Rovilos et al. 2012; Banerji et al. 2015) or not impact the star formation (e.g. Hatziminaoglou et al. 2010; Shao et al. 2010; Harrison et al. 2012; Stanley et al. 2015; Suh et al. 2017; Ramasawmy et al. 2019).

Taking this into account, in Fig. 5 the distribution of O/H abundances for our sample of wAGN and sAGN LINERs is compared with:

- Oxygen estimates for 463 Seyfert 2 galaxies ($z < 0.4$) studied by Carvalho et al. (2020). These estimates were obtained through the semi-empirical calibration:

$$(Z/Z_\odot) = a^{N2} + b, \qquad (2)$$

where $a = 4.01 \pm 0.08$ and $b = -0.07 \pm 0.01$. The observational data used by these authors were taken from the Sloan Digital Sky Survey DR7 (SDSS, York et al. 2000) and the objects have masses in the range of $9.4 \lesssim \log(M_*/M_\odot) \lesssim 11.3$.

- O/H values for 43 LINERs ionized by post-AGB stars and estimated through the semi-empirical calibration proposed by Oliveira et al. (2022):

$$12 + \log(\text{O/H}) = 0.71(\pm0.03)N2 + 8.58(\pm0.01). \quad (3)$$

The observational data used by these authors were taken from the MaNGA database, being the sample with redshift in the range of $0.2 \lesssim z \lesssim 0.7$ and mass in the interval $9.9 \lesssim \log(M_*/M_\odot) \lesssim 11.2$.

The range of galactic masses of the samples by Carvalho et al. (2020), Oliveira et al. (2022) and in the present work are similar, not yielding any bias in the $Z$ comparison. In Fig. 5, we compare the O/H estimate distributions for these three different samples, finding a very similar behavior between them. The number of objects in each sample together with the maximum and minimum (range) and the average oxygen abundance values are listed in Table 1. This result could suggest that these distinct object classes could have a similar star formation history and the recent star formation found in galaxies hosting an AGN or a LINER seems to not alter their metallicity. This result is consonance with the one derived by Stasińska et al. (2015), who based on spectral synthesis results of a large sample of SDSS objects, found that the specific star formation rates (sSFRs) in retired galaxies are identical to those of SF and AGN galaxies.

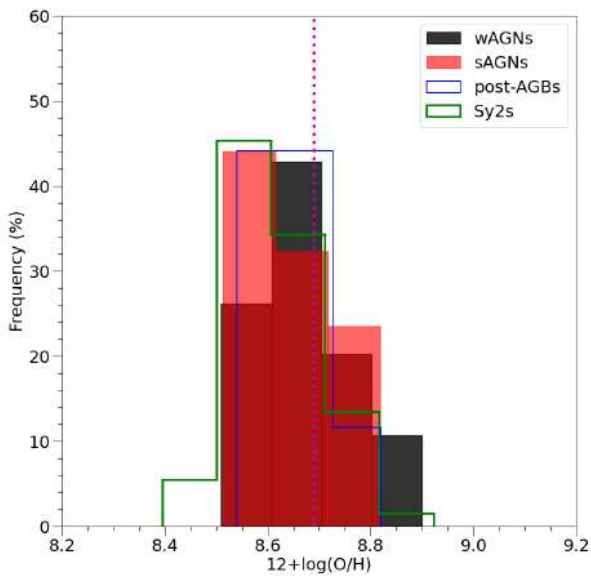

**Figure 5.** Histograms representing the oxygen abundance distributions derived for wAGN and sAGN LINERs (in black and red, respectively) by using Eq. 1, for post-AGB LINERs (in blue) from Oliveira et al. (2022) and for Seyfert 2 galaxies (in green) from Carvalho et al. (2020), as indicated. The dashed pink line represents the solar value of $12 + \log(\text{O/H})_\odot = 8.69$ (Grevesse et al., 2010).

**Table 1.** Statistics of the oxygen abundances for LINERs and Seyfert 2 nuclei. Second column indicates the number of objects in each sample.

| Object class | N | 12+log(O/H) Range | 12+log(O/H) Mean | Reference |
|---|---|---|---|---|
| wAGN LINERs | 84 | 8.50, 8.90 | 8.68 ± 0.09 | This work |
| sAGN LINERs | 34 | 8.51, 8.81 | 8.65 ± 0.09 | This work |
| post-AGB LINERs | 43 | 8.54, 8.82 | 8.65 ± 0.06 | Oliveira et al. (2022) |
| Seyfert 2 | 463 | 8.39, 8.92 | 8.61 ± 0.08 | Carvalho et al. (2020) |

## 4.2 Relation between ionization parameter and the equivalent width of Hα

The ionization parameter is defined, basically, by the ratio between the hydrogen ionizing photon flux, $Q(\text{H})$, and the density of hydrogen atoms (e.g. Dopita & Sutherland 2003; Osterbrock & Ferland 2006). For SFs, $Q(\text{H})$ is driven by the effective temperature ($T_{eff}$) of the hottest ionizing stars (which decreases with the age of main sequence stars) as demonstrated by Dors et al. (2017b). Due to the effects of opacity and/or line-blanketing in stellar atmospheres, stars with higher $Z$ tend to present lower $T_{eff}$ than those with lower $Z$ and similar mass (see Zinchenko et al. 2019a and references therein). In this sense, if gas-embedded ionizing stars and the SF gas phase have similar $Z$ (e.g. Toribio San Cipriano et al. 2017), it is expected an anti-correlation between $U$ and the metallicity of the gas phase. However, such correlation/anti-correlation for SFs is still under discussion in the literature, with studies finding correlations (e.g. Dopita et al. 2006; Morisset et al. 2016; Ji & Yan 2022; Espinosa-Ponce et al. 2022) or not (e.g. Dors et al. 2011; Poetrodjojo et al. 2018; Kreckel et al. 2019; Kumari 2021). These controversial results are indicative that $U$ also depends on other physical parameters (e.g. the nebular geometry).

In fact, if $U$ is driven, mainly, by the hardness of the ionizing radiation flux (e.g. Steidel et al. 2014), it would be derived higher $U$ values in AGN (which have harder far-UV spectra than stellar populations) in comparison to estimates for SFs. Nevertheless, similar $U$ values between SFs and AGN were found, for instance, by Pérez-Montero et al. (2019), who carried out an analysis through a comparison between results from the HCm code (Pérez-Montero, 2014) and optical spectroscopic data.

With the goal to compare our LINER $U$ values with those of AGN, in Fig. 6, the wAGN and sAGN LINERs $U$ estimates are plotted with those derived for post-AGB LINERs derived by Oliveira et al. (2022) and those for Seyfert 2 nuclei by Carvalho et al. (2020). All $U$ and O/H estimates are derived from a comparison between results obtained from CLOUDY photoionization models and observational data by using the [O III]/[O II] versus $N2$ diagram. We can see in Fig. 6 that the wAGN, sAGN, and post-AGB LINERs have similar and a narrow range of $U$ values, i.e. $-3.2 \lesssim \log U \lesssim -3.8$. This result suggests that the (possible) distinct ionizing source of both LINER types does not alter the $U$ values. Moreover, a clear tendency of LINERs to present lower $U$ values than Seyfert 2s (although sharing the bottom part of the $U$ distribution), as suggested by Ferland & Netzer (1983), is noted in Fig. 6.

It is interesting to analyze how the equivalent width of Hα depends on the ionization parameter of LINERs. The $EW_{Hα}$ is calculated as the ratio between the Hα emission-line flux, which is proportional to $Q(H)$, and its surrounding continuum fluxes (Dottori, 1981). For SFs the decrease of $EW_{Hα}$ is mainly due to the decrease of the $T_{eff}$ of O/B stars (or increase of the stellar ionizing cluster age, e.g. Copetti et al. 1985; Stasińska & Leitherer 1996; Dors et al. 2008) and becoming significant the contribution from longer-lived, non-ionizing, lower-mass stars with aging (e.g. Fernandes et al. 2003). For AGN, the decrement of the equivalent widths is mainly due to the ionizing continuum softening with increasing the luminosity $L$ (e.g. Binette et al. 1989; Netzer et al. 1992; Korista et al. 1998), the so-called "Baldwin effect" (Baldwin, 1977). The Baldwin effect (see also Shields 2007 for a review) has been found by using equivalent widths of broad (e.g. Dietrich et al. 2002; Wang et al. 2022) and narrow (e.g. Zhang et al. 2013) emission lines. Thus, since it is assumed that post-AGB LINERs and those classified as sAGN and wAGN LINERs have distinct ionizing sources, it is expected that different LINER types follow distinct $U$-$EW_{Hα}$ relation.

In Fig. 7, a plot of $\log U$ versus $\log(EW_{Hα})$, our estimates (considering the average values of $\log U$ estimated from the upper and middle diagrams in Fig. 3) for wAGN and sAGN LINERs and those for post-AGB LINERs from Oliveira et al. (2022) are shown. Assuming bins of 0.15 dex in $\log(EW_{Hα})$, we derived average $\log(EW_{Hα})$ and $\log U$ values for each bin. The black points and their error bars in Fig. 7 represents these average values and their corresponding standard deviations. Despite the scattering of the points, it can be seen a decrement of $\log U$ with $\log(EW_{Hα})$. A linear regression to the average values in these bins, whith a correlation coefficient of $r = -0.94$ and a p-value of 0.000087, is given by:

$$\log U = (-0.22 \pm 0.01) \times \log[EW_{Hα}] - 3.38 \pm 0.02. \quad (4)$$

This result could be due to two factors:

1. Post-AGB stars tend to have a harder ionizing photon flux [or $Q(H)$] than wAGN and sAGN LINERs.
2. Post-AGB stars are spread along the nebulae, maintaining a high ionizing degree at large distances (e.g. at scales of kpc), as found by Krabbe et al. (2021). Otherwise, in wAGN and sAGN LINERs the ionizing source, i.e the AGN-like source, extends to ∼ 10 pc (e.g. Constantin et al. 2015), yielding lower ionization levels as the distance increases with the nebular radius.

### 4.3 Mass metallicity relation

Finally, we discuss the mass-metallicity relation of galaxies (Lequeux et al., 1979) by using our estimates. It is well known the existence of a strong correlation between the mass and metallicity (mass metallicity relation – MZR) in elliptical and spiral bulges (e.g. Zaritsky et al. 1994; Pérez-Montero et al. 2006; Duarte Puertas et al. 2022). This relation is poorly known in AGN. In fact, Dors et al. (2014) and Nagao et

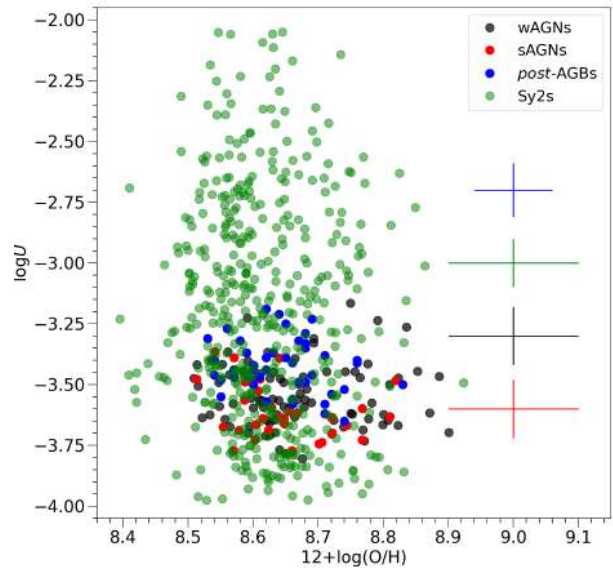

**Figure 6.** Logarithm of the ionization parameter versus oxygen abundance. Black and red points represent wAGN and sAGN LINERs, respectively, analyzed in the present work. Blue points represent estimates of post-AGB LINERs taken from Oliveira et al. (2022). Green points are estimates derived by Carvalho et al. (2020) for Seyfert 2 nuclei.

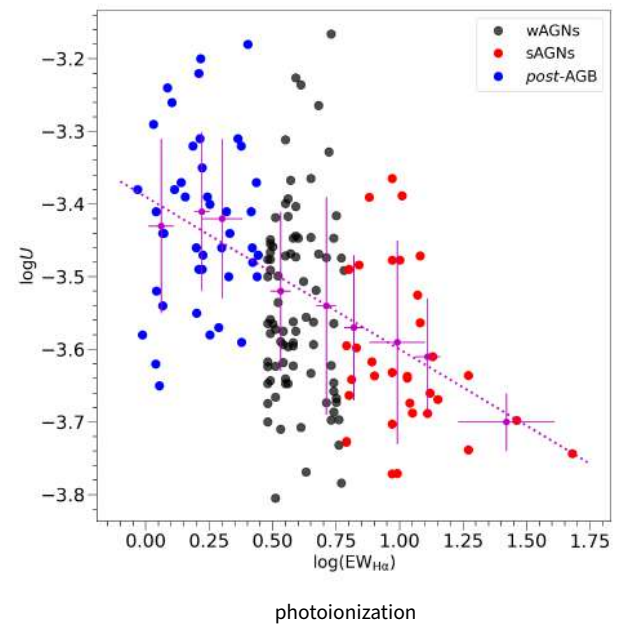

photoionization

**Figure 7.** $\log U$ versus $\log(EW_{H_I})$. Black and red points represent the average values of $\log U$ derived from interpolation between observational data and the results of photoionization model grids for wAGN and sAGN LINERs, respectively. Blue points represent values reported by Oliveira et al. (2022) for post-AGB LINERs. Pink points and error bars represent the average values and standard deviation of the points, respectively, considering bins of $\log(EW_{H_I})$ equal to 0.15 dex. The line represents the linear regression (Eq. 4) to the pink points.

(2006) found a small increase of metallicity for Seyfet 2, quasar and radio galaxies. Thomas et al. (2019) analyzed the MZR in a large sample of local AGN, comparing observational data from the SDSS and a four-dimensional grid of photoionization models using the Bayesian parameter code NEBULABAYES. These authors found that the oxygen abundance of AGN increases by ∼ 0.1 dex as a function of the stellar mass of the hosting galaxy. None of these studies analyzed the MZR for LINER nuclei. In view of that, we analyzed if the oxygen abundances of the central zone of our LINER galaxies are correlated with the stellar mass of the hosting galaxies for our sample. The stellar masses of the galaxies in our sample were taken from the database provide by Sánchez et al. (2016), and are in the range of $9.0 \lesssim \log(M_*/M_\odot) \lesssim 11.2$. In Fig. 8, our O/H estimates are plotted against the stellar masses (in units of the solar mass) of the galaxies in our sample, as well as the estimations of the oxygen abundance versus stellar masses for a sample of galaxies with SF nuclei taken from the MaNGA survey. For these SF objects, we applied the R calibration proposed by Pilyugin & Grebel (2016) to derive the oxygen abundance. Since we are comparing two different samples of objects: LINERs and SFs, obviously we derived the metallicity by using two different calibrations, one for each kind of object. Thus, some bias could be introduced in this analysis. From these estimates, we found a Pearson's correlation coefficient for our LINERs of $r = 0.24$ and a p-value of $0.008$. These coefficients indicate that the relationship between these two parameters (mass and metallicity), if any, is small.

Small correlation between mass and metallicity was also found by Pérez-Díaz et al. (2021) and Li et al. (2024). Pérez-Díaz et al. (2021) analyzed a sample of SF galaxies, Seyfert nuclei, and LINER nuclei and could derive an MZR for SF objects, while no significant correlations were found for Seyfert nor LINER nuclei. Recently, Li et al. (2024) applied the NEBULABAYES code to a sample of objects taken from the MaNGA survey and studied the MZR in active and non-active galaxies. These authors found that for galaxies that show no evidence of AGN, the $Z$ increases with $M_*$ below $M_* \sim 10^{10.5}$ M$_\odot$ and flattens at higher masses. Galaxies hosting AGN (Seyferts, Composite, Ambiguous, and LINERs) present similar O/H to non-AGN galaxies at stellar masses above $10^{10.5}$ M$_\odot$, biasing to higher O/H below this stellar mass. However, it is important to note that for a reliable statistical result, a larger sample of LINER galaxies should be analyzed. Also, it is worth mentioning that the values we derive for the oxygen abundance are in the high-metallicity end and the metallicity range is not very large, hence it is difficult to clearly establish the relationship between mass and metallicity in these galaxies.

## 5. Conclusions

We compiled the optical spectroscopic data of 118 galaxies taken from the MaNGA survey classified as LINERs using three different BPT diagrams and sub-classified as weak and strong AGN (84 and 34 objects, respectively) from the WHAN diagnostic diagram. Comparing observational data with photoionization model grids built with the CLOUDY code we derive

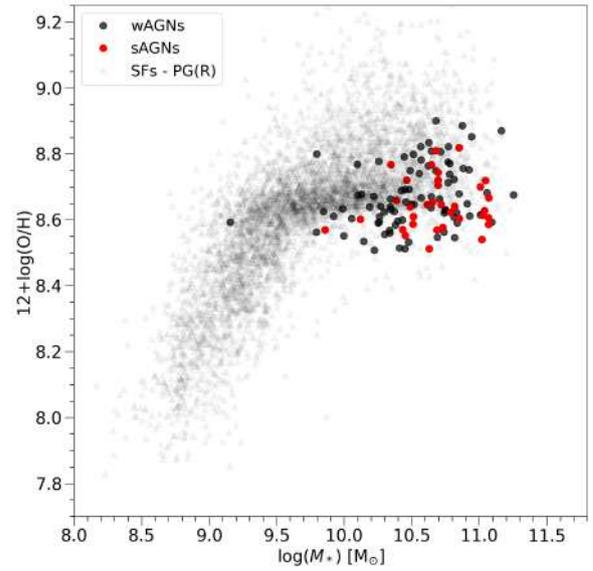

**Figure 8.** Oxygen abundances derived through our $Z - N2$ calibration (Eq. 1) versus stellar masses (in units of the solar mass). Black and red points represent wAGN and sAGN LINERs, respectively. Black triangles are oxygen abundances of SF nuclei derived by using the calibration proposed by Pilyugin & Grebel (2016)

a semi-empirical calibration based on the $N2$ index to estimate the oxygen abundances for these objects. Through our calibration, we derived oxygen abundances for our wAGN LINERs in the range $8.50 \lesssim 12 + \log(O/H) \lesssim 8.90$, with an average value of $12 + \log(O/H) = 8.68$, and for our sAGN LINERs in the range $8.51 \lesssim 12 + \log(O/H) \lesssim 8.81$, with an average value of $12 + \log(O/H) = 8.65$. Both, wAGN and sAGN LINERs, present very similar metallicity ranges and average oxygen abundance values, and about 30 per cent of them have oversolar abundances.

The O/H abundances derived through the calibration proposed in this work are in consonance with those derived by using calibrations for a sample of 463 Sy2 galaxies, as well as a calibration for a sample of 43 LINER galaxies ionized by hot post-AGB stars and classified as retired galaxies in the WHAN diagram. In fact, we found a very good agreement between O/H derived by using the calibration proposed for LINERs classified as retired galaxies and the O/H obtained for our wAGN and sAGN LINERs, suggesting that both calibrations can be applied for this kind of nuclei.

We also search for the existence of a relation between the equivalent width of the observed Hα emission line and the estimated ionization parameter given by the photoionization models, deriving a semi-empirical linear relation between them. We finally studied the mass-metallicity relationship for this kind of galaxies. Our data show that the correlation between mass and metallicity, if any, is small, but this is probably biased as we are considering the metallicity of the center of the galaxy as representative (or something similar).

## Data Availability
Not applicable

## Acknowledgments
We thanks a lot the referee, Dr. Ángel R. López-Sánchez, for all comments and suggestions, which improved a lot our paper. CBO is grateful to the Fundação de Amparo à Pesquisa do Estado de São Paulo (FAPESP) for the support under grant 2023/10182–0 and to the Coordenação de Aperfeiçoamento de Pessoal de Nível Superior (CAPES). OLD is grateful to FAPESP, process number 2022/07066-6, and to Conselho Nacional de Desenvolvimento Científico e Tecnológico (CNPq).

## Appendix 1.  Fluxes of emission lines

**Table 2.** Reddening corrected emission-line intensities (in relation to Hβ=1.00) derived for each LINER nucleus in our sample. Values of the logarithm of EW$_{Hα}$, c(Hβ), and the absolute flux of Hβ in units of $10^{-17}$ erg/s/cm²/spaxel are also listed.

| Plate-IFU | [O II] λ3727 | [O III] λ5007 | [OI] λ6300 | Hα | [N II] λ6584 | [S II] λ6717 | [S II] λ6731 | log(EW$_{Hα}$) | c(Hβ) | F(Hβ) |
|---|---|---|---|---|---|---|---|---|---|---|
| 8086-12705 | 7.81 ± 0.18 | 3.43 ± 0.06 | 1.42 ± 0.08 | 2.87 ± 0.06 | 5.44 ± 0.06 | 2.45 ± 0.10 | 1.60 ± 0.09 | 0.65 | 0.81 | 10.71 ± 0.71 |
| 10510-6103 | 20.02 ± 0.47 | 2.88 ± 0.08 | 1.23 ± 0.05 | 2.83 ± 0.06 | 4.43 ± 0.07 | 1.29 ± 0.03 | 1.43 ± 0.04 | 0.79 | 1.58 | 190.30 ± 18.53 |
| 9193-9101 | 3.24 ± 0.06 | 2.55 ± 0.07 | 0.96 ± 0.05 | 2.63 ± 0.18 | 4.81 ± 0.19 | 1.50 ± 0.11 | 1.09 ± 0.10 | 0.68 | 0.01 | 3.65 ± 0.17 |
| 8482-12704 | 9.87 ± 0.10 | 2.48 ± 0.05 | 1.11 ± 0.07 | 2.88 ± 0.04 | 4.01 ± 0.05 | 1.76 ± 0.05 | 1.33 ± 0.05 | 0.90 | 0.51 | 15.46 ± 0.56 |
| 11951-12704 | 5.40 ± 0.19 | 2.27 ± 0.11 | 1.16 ± 0.09 | 2.88 ± 0.14 | 5.86 ± 0.13 | 1.86 ± 0.13 | 1.20 ± 0.13 | 0.49 | 0.24 | 2.25 ± 0.24 |
| 12685-1901 | 2.70 ± 0.18 | 2.09 ± 0.13 | 1.12 ± 0.20 | 2.57 ± 0.10 | 2.43 ± 0.13 | 0.70 ± 0.10 | 1.05 ± 0.16 | 0.59 | 0.01 | 1.11 ± 0.17 |
| 9866-3701 | 4.59 ± 0.22 | 2.08 ± 0.10 | 1.32 ± 0.13 | 2.87 ± 0.08 | 2.17 ± 0.08 | 1.06 ± 0.06 | 0.81 ± 0.07 | 0.55 | 0.52 | 2.38 ± 0.34 |
| 9037-12705 | 4.48 ± 0.16 | 1.91 ± 0.12 | 1.26 ± 0.16 | 2.88 ± 0.23 | 1.95 ± 0.21 | 2.03 ± 0.23 | 2.03 ± 0.23 | 0.51 | 0.38 | 3.00 ± 0.27 |
| 8625-12704 | 3.97 ± 0.05 | 1.02 ± 0.09 | 1.32 ± 0.16 | 2.87 ± 0.07 | 5.67 ± 0.07 | 1.64 ± 0.06 | 1.42 ± 0.06 | 0.48 | 0.27 | 10.70 ± 0.38 |
| 11751-12704 | 5.03 ± 0.07 | 1.79 ± 0.04 | 1.20 ± 0.04 | 2.87 ± 0.07 | 2.51 ± 0.07 | 1.36 ± 0.02 | 1.36 ± 0.02 | 1.08 | 0.29 | 31.46 ± 1.11 |
| 8314-1902 | 5.18 ± 0.11 | 1.83 ± 0.04 | 1.08 ± 0.04 | 2.61 ± 0.07 | 1.73 ± 0.07 | 1.47 ± 0.04 | 1.10 ± 0.04 | 0.77 | 0.01 | 5.70 ± 0.16 |
| 8450-6104 | 8.36 ± 0.12 | 1.61 ± 0.04 | 1.16 ± 0.06 | 2.87 ± 0.07 | 6.02 ± 0.08 | 2.39 ± 0.09 | 1.91 ± 0.10 | 0.76 | 0.48 | 13.64 ± 0.58 |
| 11834-9101 | 5.93 ± 0.11 | 1.53 ± 0.06 | 1.16 ± 0.08 | 2.87 ± 0.06 | 4.31 ± 0.07 | 2.03 ± 0.05 | 1.41 ± 0.05 | 0.54 | 0.32 | 4.29 ± 0.24 |
| 11947-12704 | 7.61 ± 0.10 | 1.28 ± 0.02 | 1.13 ± 0.05 | 2.86 ± 0.07 | 4.00 ± 0.07 | 1.90 ± 0.13 | 1.91 ± 0.14 | 1.46 | 0.59 | 70.96 ± 1.44 |
| 8652-9102 | 10.74 ± 0.15 | 1.81 ± 0.03 | 1.09 ± 0.03 | 2.87 ± 0.06 | 4.24 ± 0.06 | 1.75 ± 0.04 | 1.96 ± 0.05 | 1.15 | 1.02 | 35.14 ± 0.98 |
| 8551-3704 | 8.00 ± 0.22 | 1.67 ± 0.09 | 1.09 ± 0.09 | 2.89 ± 0.08 | 4.96 ± 0.08 | 2.09 ± 0.26 | 2.17 ± 0.28 | 0.56 | 0.66 | 7.58 ± 0.71 |
| 12066-6102 | 4.47 ± 0.12 | 1.56 ± 0.09 | 1.08 ± 0.10 | 2.88 ± 0.12 | 4.97 ± 0.12 | 1.91 ± 0.10 | 1.68 ± 0.11 | 0.48 | 0.19 | 1.35 ± 0.17 |
| 12487-6104 | 8.03 ± 0.13 | 1.65 ± 0.05 | 1.06 ± 0.07 | 2.87 ± 0.06 | 4.60 ± 0.06 | 2.17 ± 0.04 | 1.73 ± 0.04 | 0.55 | 0.41 | 11.77 ± 0.39 |
| 10511-12702 | 3.87 ± 0.15 | 0.98 ± 0.19 | 1.09 ± 0.10 | 2.88 ± 0.13 | 2.94 ± 0.13 | 1.55 ± 0.16 | 1.28 ± 0.21 | 0.50 | 0.11 | 0.77 ± 0.08 |
| 7972-6103 | 6.74 ± 0.07 | 1.03 ± 0.02 | 1.04 ± 0.04 | 2.87 ± 0.04 | 2.59 ± 0.04 | 2.04 ± 0.06 | 1.33 ± 0.05 | 1.05 | 0.46 | 25.27 ± 0.67 |
| 8562-6102 | 2.30 ± 0.11 | 1.02 ± 0.05 | 1.04 ± 0.17 | 2.87 ± 0.06 | 2.72 ± 0.06 | 1.36 ± 0.06 | 0.84 ± 0.05 | 0.59 | 0.16 | 3.49 ± 0.55 |
| 9502-9101 | 7.20 ± 0.10 | 1.16 ± 0.05 | 1.00 ± 0.07 | 2.89 ± 0.05 | 2.76 ± 0.05 | 1.80 ± 0.07 | 1.06 ± 0.06 | 0.75 | 0.57 | 7.56 ± 0.36 |
| 8158-3703 | 9.13 ± 0.21 | 1.72 ± 0.02 | 1.01 ± 0.03 | 2.87 ± 0.02 | 2.03 ± 0.02 | 3.22 ± 0.03 | 2.42 ± 0.02 | 3.21 | 0.19 | 4.64 ± 0.07 |
| 9044-12701 | 6.81 ± 0.19 | 1.79 ± 0.07 | 1.00 ± 0.09 | 2.86 ± 0.06 | 3.67 ± 0.06 | 2.00 ± 0.05 | 1.69 ± 0.06 | 0.54 | 0.69 | 8.92 ± 0.54 |
| 8597-12702 | 7.35 ± 0.14 | 1.99 ± 0.07 | 1.00 ± 0.06 | 2.87 ± 0.10 | 4.66 ± 0.11 | 2.20 ± 0.06 | 1.56 ± 0.06 | 0.59 | 0.40 | 4.83 ± 0.28 |
| 8606-12701 | 8.56 ± 0.17 | 3.89 ± 0.11 | 1.00 ± 0.06 | 2.88 ± 0.11 | 4.60 ± 0.11 | 1.54 ± 0.06 | 1.17 ± 0.06 | 0.75 | 0.56 | 6.31 ± 0.34 |
| 8133-12704 | 6.56 ± 0.10 | 1.22 ± 0.05 | 0.98 ± 0.03 | 2.91 ± 0.06 | 2.98 ± 0.07 | 1.76 ± 0.05 | 1.26 ± 0.05 | 1.03 | 0.48 | 2.70 ± 0.09 |
| 12518-6102 | 4.34 ± 0.08 | 1.45 ± 0.05 | 0.97 ± 0.06 | 2.88 ± 0.06 | 2.29 ± 0.06 | 1.15 ± 0.07 | 1.14 ± 0.08 | 0.78 | 0.32 | 4.88 ± 0.41 |
| 8454-6102 | 6.55 ± 0.12 | 1.58 ± 0.17 | 0.98 ± 0.06 | 2.90 ± 0.08 | 4.38 ± 0.07 | 1.76 ± 0.09 | 1.31 ± 0.09 | 0.73 | 0.38 | 2.99 ± 0.20 |
| 9091-6101 | 14.48 ± 0.22 | 1.59 ± 0.07 | 0.95 ± 0.09 | 2.86 ± 0.08 | 3.18 ± 0.08 | 1.54 ± 0.10 | 1.58 ± 0.12 | 0.63 | 0.92 | 26.37 ± 1.62 |
| 8995-3703 | 3.50 ± 0.10 | 1.76 ± 0.06 | 0.94 ± 0.06 | 2.88 ± 0.06 | 2.43 ± 0.06 | 1.43 ± 0.07 | 1.29 ± 0.08 | 0.65 | 0.10 | 2.54 ± 0.17 |
| 8937-1902 | 4.89 ± 0.11 | 2.09 ± 0.05 | 0.96 ± 0.06 | 2.87 ± 0.06 | 2.33 ± 0.06 | 0.91 ± 0.04 | 1.08 ± 0.05 | 0.56 | 0.42 | 10.15 ± 0.48 |
| 11745-6104 | 5.16 ± 0.20 | 4.29 ± 0.08 | 0.94 ± 0.06 | 2.87 ± 0.08 | 4.75 ± 0.10 | 1.28 ± 0.04 | 1.32 ± 0.04 | 0.61 | 0.59 | 7.40 ± 0.39 |
| 10842-6102 | 3.91 ± 0.12 | 3.95 ± 0.10 | 0.91 ± 0.15 | 2.87 ± 0.11 | 4.30 ± 0.12 | 1.21 ± 0.08 | 1.21 ± 0.09 | 0.73 | 0.50 | 8.18 ± 0.41 |
| 10498-1902 | 7.87 ± 0.22 | 3.19 ± 0.12 | 0.91 ± 0.12 | 2.89 ± 0.17 | 4.42 ± 0.16 | 2.98 ± 0.30 | 2.98 ± 0.30 | 0.60 | 0.53 | 3.69 ± 0.43 |
| 8255-6101 | 7.77 ± 0.12 | 2.71 ± 0.04 | 0.92 ± 0.04 | 2.88 ± 0.04 | 2.90 ± 0.04 | 1.36 ± 0.04 | 1.08 ± 0.04 | 0.97 | 0.80 | 16.80 ± 0.74 |
| 8978-9101 | 7.27 ± 0.14 | 1.84 ± 0.05 | 0.92 ± 0.06 | 2.86 ± 0.08 | 4.49 ± 0.09 | 2.18 ± 0.05 | 1.65 ± 0.04 | 0.83 | 0.55 | 11.41 ± 0.43 |
| 8445-12702 | 5.74 ± 0.08 | 1.82 ± 0.06 | 0.92 ± 0.08 | 2.87 ± 0.07 | 1.97 ± 0.07 | 1.47 ± 0.07 | 1.61 ± 0.08 | 0.62 | 0.19 | 7.85 ± 0.41 |
| 8616-1902 | 5.00 ± 0.08 | 1.75 ± 0.25 | 0.91 ± 0.07 | 2.87 ± 0.06 | 3.20 ± 0.06 | 1.69 ± 0.04 | 1.17 ± 0.04 | 0.49 | 0.27 | 6.71 ± 0.36 |
| 8147-6102 | 8.14 ± 0.19 | 1.60 ± 0.05 | 0.93 ± 0.05 | 2.88 ± 0.11 | 5.12 ± 0.11 | 1.90 ± 0.08 | 1.88 ± 0.09 | 0.75 | 0.68 | 4.40 ± 0.22 |
| 8080-12703 | 8.29 ± 0.25 | 1.58 ± 0.06 | 0.89 ± 0.08 | 2.86 ± 0.09 | 4.82 ± 0.09 | 1.57 ± 0.05 | 1.16 ± 0.05 | 0.51 | 0.78 | 16.79 ± 0.87 |
| 12667-12704 | 4.67 ± 0.09 | 1.42 ± 0.07 | 0.93 ± 0.08 | 2.88 ± 0.08 | 3.63 ± 0.09 | 1.74 ± 0.07 | 1.29 ± 0.07 | 0.67 | 0.16 | 2.23 ± 0.16 |
| 9881-1902 | 2.76 ± 0.13 | 1.04 ± 0.06 | 0.83 ± 0.09 | 2.55 ± 0.19 | 2.59 ± 0.26 | 0.95 ± 0.06 | 0.92 ± 0.06 | 0.49 | 0.01 | 1.77 ± 0.12 |
| 8597-3703 | 5.22 ± 0.11 | 0.97 ± 0.04 | 0.93 ± 0.05 | 2.87 ± 0.08 | 3.26 ± 0.08 | 1.64 ± 0.06 | 1.82 ± 0.07 | 0.81 | 0.29 | 15.16 ± 0.61 |
| 11761-12705 | 7.18 ± 0.15 | 1.09 ± 0.03 | 0.88 ± 0.04 | 2.85 ± 0.08 | 3.01 ± 0.08 | 1.79 ± 0.07 | 1.45 ± 0.08 | 1.11 | 1.03 | 40.61 ± 1.55 |
| 8318-6102 | 9.75 ± 0.12 | 1.37 ± 0.03 | 0.89 ± 0.03 | 2.93 ± 0.11 | 3.87 ± 0.12 | 2.48 ± 0.12 | 0.58 ± 0.11 | 1.68 | 0.79 | 32.34 ± 0.75 |
| 9486-9101 | 6.83 ± 0.23 | 1.75 ± 0.09 | 0.86 ± 0.16 | 2.87 ± 0.12 | 2.88 ± 0.11 | 1.70 ± 0.13 | 0.71 ± 0.10 | 0.58 | 0.88 | 9.25 ± 0.85 |
| 8483-12703 | 7.22 ± 0.11 | 3.35 ± 0.09 | 0.85 ± 0.05 | 2.88 ± 0.06 | 3.20 ± 0.05 | 1.82 ± 0.05 | 1.22 ± 0.04 | 0.88 | 0.47 | 4.16 ± 0.21 |
| 11830-12701 | 9.85 ± 0.32 | 1.45 ± 0.08 | 0.85 ± 0.06 | 2.86 ± 0.07 | 3.11 ± 0.07 | 1.32 ± 0.06 | 0.91 ± 0.05 | 0.73 | 0.88 | 7.29 ± 0.54 |
| 11965-9102 | 9.86 ± 0.51 | 1.43 ± 0.08 | 0.84 ± 0.09 | 2.85 ± 0.06 | 2.71 ± 0.06 | 1.39 ± 0.08 | 1.11 ± 0.09 | 0.48 | 1.14 | 9.29 ± 1.01 |
| 10837-9102 | 5.61 ± 0.09 | 1.87 ± 0.05 | 0.85 ± 0.03 | 2.87 ± 0.04 | 2.68 ± 0.04 | 1.68 ± 0.05 | 1.34 ± 0.05 | 0.80 | 0.32 | 9.93 ± 0.34 |
| 8333-12701 | 4.07 ± 0.06 | 1.20 ± 0.04 | 0.82 ± 0.04 | 2.88 ± 0.06 | 2.88 ± 0.06 | 1.26 ± 0.05 | 1.02 ± 0.05 | 1.07 | 0.09 | 3.63 ± 0.12 |
| 8656-6103 | 7.07 ± 0.10 | 2.79 ± 0.05 | 0.82 ± 0.11 | 2.88 ± 0.06 | 3.93 ± 0.07 | 1.63 ± 0.04 | 1.33 ± 0.04 | 0.74 | 0.52 | 6.61 ± 0.24 |

| Plate-IFU | [O II] λ3727 | [O III] λ5007 | [O I] λ6300 | Hα | [N II] λ6584 | [S II] λ6717 | [S II] λ6731 | log(EW$_{\mathrm{H}\alpha}$) | c(Hβ) | F(Hβ) |
|---|---|---|---|---|---|---|---|---|---|---|
| 8981-6101 | 4.86 ± 0.12 | 2.95 ± 0.07 | 0.81 ± 0.06 | 2.87 ± 0.05 | 3.72 ± 0.06 | 1.32 ± 0.05 | 1.18 ± 0.05 | 0.55 | 0.45 | 6.36 ± 0.32 |
| 8988-6102 | 13.35 ± 0.28 | 1.44 ± 0.08 | 0.79 ± 0.05 | 2.89 ± 0.05 | 3.31 ± 0.05 | 1.33 ± 0.10 | 1.13 ± 0.11 | 0.77 | 1.46 | 28.46 ± 2.12 |
| 11746-6102 | 4.50 ± 0.12 | 1.46 ± 0.08 | 0.78 ± 0.12 | 2.88 ± 0.12 | 3.50 ± 0.13 | 1.91 ± 0.26 | 1.95 ± 0.31 | 0.52 | 0.18 | 1.98 ± 0.17 |
| 9040-9102 | 4.86 ± 0.16 | 1.24 ± 0.06 | 0.79 ± 0.07 | 2.88 ± 0.07 | 3.47 ± 0.08 | 1.25 ± 0.04 | 0.78 ± 0.03 | 0.74 | 0.69 | 6.46 ± 0.35 |
| 8329-3701 | 4.84 ± 0.06 | 1.23 ± 0.04 | 0.79 ± 0.03 | 2.89 ± 0.04 | 2.68 ± 0.04 | 1.25 ± 0.04 | 0.88 ± 0.04 | 1.08 | 0.36 | 5.88 ± 0.19 |
| 8602-12701 | 9.48 ± 0.11 | 2.10 ± 0.03 | 0.76 ± 0.03 | 2.86 ± 0.06 | 4.95 ± 0.07 | 1.48 ± 0.06 | 1.15 ± 0.05 | 0.97 | 0.63 | 19.00 ± 0.52 |
| 7990-12704 | 4.73 ± 0.17 | 1.66 ± 0.09 | 0.75 ± 0.09 | 2.87 ± 0.08 | 2.73 ± 0.07 | 1.78 ± 0.06 | 1.26 ± 0.06 | 0.48 | 0.35 | 3.01 ± 0.18 |
| 10843-1901 | 12.81 ± 0.45 | 1.80 ± 0.09 | 0.74 ± 0.06 | 2.86 ± 0.08 | 3.03 ± 0.08 | 1.55 ± 0.05 | 1.07 ± 0.05 | 0.61 | 1.00 | 11.27 ± 0.85 |
| 8562-9102 | 5.65 ± 0.24 | 1.88 ± 0.09 | 0.74 ± 0.07 | 2.87 ± 0.07 | 3.25 ± 0.07 | 1.08 ± 0.05 | 0.70 ± 0.04 | 0.49 | 0.61 | 6.97 ± 0.45 |
| 9888-12701 | 7.02 ± 0.11 | 2.62 ± 0.06 | 0.74 ± 0.06 | 2.87 ± 0.10 | 5.06 ± 0.10 | 2.14 ± 0.16 | 1.55 ± 0.17 | 0.84 | 0.52 | 16.50 ± 0.55 |
| 12067-12702 | 7.67 ± 0.36 | 2.95 ± 0.20 | 0.73 ± 0.12 | 2.89 ± 0.17 | 3.12 ± 0.16 | 1.10 ± 0.13 | 0.95 ± 0.13 | 0.49 | 0.60 | 2.23 ± 0.37 |
| 10221-6104 | 7.65 ± 0.17 | 1.29 ± 0.08 | 0.73 ± 0.05 | 2.86 ± 0.05 | 4.49 ± 0.05 | 1.96 ± 0.09 | 1.26 ± 0.08 | 0.74 | 0.69 | 10.76 ± 0.66 |
| 10507-6104 | 7.74 ± 0.14 | 1.31 ± 0.07 | 0.73 ± 0.08 | 2.87 ± 0.07 | 4.19 ± 0.07 | 2.20 ± 0.08 | 1.52 ± 0.08 | 0.48 | 0.35 | 9.19 ± 0.52 |
| 8591-6101 | 5.90 ± 0.16 | 1.07 ± 0.12 | 0.74 ± 0.10 | 2.89 ± 0.11 | 3.03 ± 0.10 | 1.97 ± 0.18 | 1.29 ± 0.21 | 0.48 | 0.53 | 3.71 ± 0.38 |
| 8593-3703 | 7.56 ± 0.12 | 0.89 ± 0.03 | 0.73 ± 0.03 | 2.87 ± 0.08 | 3.38 ± 0.08 | 1.43 ± 0.04 | 1.41 ± 0.04 | 0.99 | 0.62 | 25.24 ± 0.57 |
| 8309-6101 | 4.93 ± 0.14 | 1.84 ± 0.08 | 0.71 ± 0.08 | 2.88 ± 0.09 | 4.48 ± 0.10 | 1.55 ± 0.20 | 1.56 ± 0.23 | 0.55 | 0.17 | 2.39 ± 0.19 |
| 11013-3702 | 6.54 ± 0.09 | 1.38 ± 0.03 | 0.70 ± 0.03 | 2.86 ± 0.04 | 3.32 ± 0.05 | 1.44 ± 0.03 | 1.03 ± 0.03 | 1.13 | 0.53 | 29.81 ± 0.57 |
| 11749-6104 | 5.86 ± 0.12 | 0.99 ± 0.04 | 0.70 ± 0.06 | 2.90 ± 0.06 | 2.87 ± 0.06 | 1.47 ± 0.07 | 0.92 ± 0.06 | 0.80 | 0.61 | 6.82 ± 0.31 |
| 8336-12704 | 11.28 ± 1.19 | 1.26 ± 0.19 | 0.68 ± 0.09 | 2.84 ± 0.09 | 3.50 ± 0.10 | 0.87 ± 0.05 | 0.68 ± 0.07 | 0.51 | 1.35 | 9.00 ± 1.37 |
| 7958-3702 | 6.87 ± 0.28 | 1.26 ± 0.08 | 0.69 ± 0.10 | 2.86 ± 0.05 | 2.37 ± 0.05 | 1.34 ± 0.08 | 0.77 ± 0.06 | 0.49 | 0.92 | 9.11 ± 0.74 |
| 8150-6104 | 5.73 ± 0.17 | 1.64 ± 0.07 | 0.68 ± 0.07 | 2.86 ± 0.06 | 3.50 ± 0.06 | 1.37 ± 0.04 | 1.19 ± 0.04 | 0.52 | 0.65 | 11.18 ± 0.56 |
| 9024-12705 | 9.97 ± 0.18 | 1.39 ± 0.04 | 0.68 ± 0.03 | 2.88 ± 0.09 | 3.86 ± 0.09 | 1.42 ± 0.04 | 0.97 ± 0.05 | 1.27 | 0.82 | 24.22 ± 0.85 |
| 11969-6102 | 4.77 ± 0.14 | 1.09 ± 0.04 | 0.68 ± 0.06 | 2.87 ± 0.06 | 3.26 ± 0.06 | 1.14 ± 0.06 | 0.94 ± 0.06 | 0.58 | 0.57 | 10.85 ± 1.42 |
| 11754-9102 | 7.40 ± 0.33 | 1.03 ± 0.15 | 0.69 ± 0.09 | 2.94 ± 0.09 | 3.05 ± 0.08 | 1.12 ± 0.11 | 0.97 ± 0.15 | 0.53 | 0.98 | 3.58 ± 0.46 |
| 11955-12703 | 5.34 ± 0.08 | 3.07 ± 0.05 | 0.65 ± 0.07 | 2.87 ± 0.07 | 3.72 ± 0.07 | 1.91 ± 0.05 | 1.35 ± 0.04 | 0.72 | 0.30 | 6.54 ± 0.25 |
| 8134-9102 | 6.68 ± 0.15 | 2.68 ± 0.12 | 0.67 ± 0.08 | 2.87 ± 0.09 | 4.22 ± 0.11 | 1.37 ± 0.10 | 2.31 ± 0.13 | 0.58 | 0.27 | 3.34 ± 0.22 |
| 11979-3703 | 7.87 ± 0.39 | 2.00 ± 0.13 | 0.66 ± 0.09 | 2.87 ± 0.12 | 3.18 ± 0.13 | 1.14 ± 0.07 | 0.94 ± 0.08 | 0.48 | 0.93 | 4.93 ± 0.47 |
| 10514-6102 | 4.79 ± 0.12 | 1.73 ± 0.14 | 0.62 ± 0.08 | 2.68 ± 0.14 | 3.32 ± 0.11 | 1.48 ± 0.25 | 1.80 ± 0.35 | 0.57 | 0.01 | 0.83 ± 0.11 |
| 8714-6102 | 3.76 ± 0.16 | 1.88 ± 0.08 | 0.64 ± 0.08 | 2.88 ± 0.06 | 3.02 ± 0.06 | 0.87 ± 0.05 | 0.78 ± 0.05 | 0.57 | 0.70 | 5.46 ± 0.35 |
| 8146-12705 | 4.42 ± 0.11 | 1.72 ± 0.06 | 0.64 ± 0.07 | 2.83 ± 0.11 | 5.04 ± 0.11 | 1.84 ± 0.07 | 1.35 ± 0.07 | 0.59 | 0.01 | 0.92 ± 0.06 |
| 9025-12701 | 7.89 ± 0.27 | 1.57 ± 0.12 | 0.64 ± 0.08 | 2.88 ± 0.12 | 2.16 ± 0.11 | 1.91 ± 0.09 | 1.35 ± 0.09 | 0.48 | 0.48 | 1.48 ± 0.15 |
| 8084-6103 | 10.24 ± 0.39 | 1.32 ± 0.07 | 0.64 ± 0.06 | 2.85 ± 0.06 | 2.80 ± 0.06 | 1.31 ± 0.05 | 0.76 ± 0.05 | 0.79 | 1.24 | 15.25 ± 0.91 |
| 9514-12702 | 4.06 ± 0.13 | 1.05 ± 0.06 | 0.65 ± 0.08 | 2.89 ± 0.08 | 3.61 ± 0.12 | 1.14 ± 0.05 | 0.77 ± 0.04 | 0.66 | 0.37 | 1.66 ± 0.11 |
| 9047-1901 | 4.09 ± 0.12 | 1.01 ± 0.05 | 0.64 ± 0.06 | 2.88 ± 0.06 | 4.05 ± 0.07 | 1.40 ± 0.04 | 1.13 ± 0.05 | 0.55 | 0.25 | 4.59 ± 0.23 |
| 8314-3704 | 4.55 ± 0.11 | 0.79 ± 0.04 | 0.63 ± 0.06 | 2.87 ± 0.06 | 3.04 ± 0.06 | 1.37 ± 0.04 | 1.04 ± 0.04 | 0.74 | 0.37 | 10.19 ± 0.35 |
| 8982-12704 | 4.83 ± 0.37 | 1.16 ± 0.11 | 0.61 ± 0.10 | 2.86 ± 0.07 | 2.49 ± 0.08 | 1.15 ± 0.07 | 0.98 ± 0.08 | 0.49 | 0.72 | 3.73 ± 0.39 |
| 7961-6101 | 6.97 ± 0.23 | 1.63 ± 0.09 | 0.61 ± 0.09 | 2.87 ± 0.12 | 3.58 ± 0.11 | 2.51 ± 0.10 | 1.62 ± 0.09 | 0.54 | 0.38 | 2.18 ± 0.20 |
| 8620-12705 | 11.20 ± 0.23 | 1.92 ± 0.04 | 0.57 ± 0.02 | 2.84 ± 0.06 | 3.23 ± 0.06 | 1.39 ± 0.05 | 1.21 ± 0.05 | 1.12 | 1.55 | 81.31 ± 3.03 |
| 9503-3704 | 6.70 ± 0.20 | 1.22 ± 0.06 | 0.57 ± 0.06 | 2.87 ± 0.04 | 3.12 ± 0.05 | 1.14 ± 0.07 | 0.93 ± 0.08 | 0.74 | 0.80 | 9.08 ± 0.51 |
| 8152-6102 | 2.84 ± 0.08 | 1.32 ± 0.04 | 0.56 ± 0.05 | 2.87 ± 0.06 | 3.43 ± 0.08 | 1.31 ± 0.06 | 0.80 ± 0.06 | 0.56 | 0.23 | 5.54 ± 0.56 |
| 8552-9102 | 3.75 ± 0.10 | 0.84 ± 0.15 | 0.55 ± 0.07 | 2.90 ± 0.10 | 2.98 ± 0.08 | 1.31 ± 0.06 | 0.83 ± 0.05 | 0.66 | 0.39 | 2.44 ± 0.17 |
| 8323-6103 | 11.13 ± 0.31 | 1.20 ± 0.06 | 0.53 ± 0.04 | 2.83 ± 0.04 | 2.47 ± 0.04 | 0.89 ± 0.03 | 0.74 ± 0.03 | 0.97 | 1.58 | 77.63 ± 5.14 |
| 8464-1902 | 4.49 ± 0.11 | 1.00 ± 0.05 | 0.53 ± 0.10 | 2.87 ± 0.06 | 2.80 ± 0.07 | 1.27 ± 0.06 | 1.29 ± 0.06 | 0.58 | 0.45 | 7.62 ± 0.31 |
| 8942-6104 | 4.84 ± 0.14 | 1.80 ± 0.06 | 0.51 ± 0.05 | 2.86 ± 0.05 | 2.44 ± 0.06 | 1.13 ± 0.03 | 0.83 ± 0.03 | 0.50 | 0.47 | 4.91 ± 0.21 |
| 8324-6104 | 10.19 ± 0.22 | 1.64 ± 0.07 | 0.50 ± 0.03 | 2.87 ± 0.06 | 4.01 ± 0.06 | 1.01 ± 0.04 | 0.59 ± 0.04 | 0.97 | 1.21 | 32.41 ± 1.78 |
| 9888-3701 | 3.67 ± 0.14 | 0.90 ± 0.05 | 0.50 ± 0.08 | 2.86 ± 0.05 | 2.89 ± 0.05 | 1.08 ± 0.10 | 0.91 ± 0.11 | 0.51 | 0.70 | 12.88 ± 0.68 |
| 9881-6102 | 3.54 ± 0.09 | 0.93 ± 0.05 | 0.48 ± 0.08 | 2.87 ± 0.06 | 2.77 ± 0.07 | 1.04 ± 0.07 | 0.85 ± 0.06 | 0.63 | 0.29 | 6.62 ± 0.51 |
| 9046-12704 | 5.55 ± 0.13 | 1.14 ± 0.07 | 0.48 ± 0.03 | 2.92 ± 0.07 | 3.52 ± 0.08 | 1.22 ± 0.07 | 0.87 ± 0.05 | 0.89 | 0.66 | 5.07 ± 0.21 |
| 11945-3704 | 6.24 ± 0.24 | 1.25 ± 0.12 | 0.47 ± 0.06 | 2.88 ± 0.13 | 3.16 ± 0.12 | 1.32 ± 0.06 | 1.15 ± 0.07 | 0.51 | 0.60 | 3.28 ± 0.30 |
| 8309-3702 | 5.54 ± 0.16 | 1.94 ± 0.06 | 0.47 ± 0.04 | 2.88 ± 0.06 | 1.96 ± 0.06 | 0.97 ± 0.03 | 0.70 ± 0.03 | 1.00 | 0.69 | 6.95 ± 0.28 |
| 8722-1901 | 4.97 ± 0.14 | 0.96 ± 0.06 | 0.47 ± 0.07 | 2.89 ± 0.06 | 3.44 ± 0.06 | 1.21 ± 0.05 | 0.87 ± 0.05 | 0.65 | 0.63 | 7.27 ± 0.39 |
| 8717-1902 | 3.37 ± 0.13 | 1.70 ± 0.07 | 0.48 ± 0.05 | 2.93 ± 0.06 | 2.27 ± 0.06 | 0.98 ± 0.07 | 0.74 ± 0.08 | 0.97 | 0.71 | 4.07 ± 0.78 |
| 9887-3704 | 3.40 ± 0.17 | 0.88 ± 0.08 | 0.44 ± 0.07 | 2.89 ± 0.08 | 3.10 ± 0.09 | 0.94 ± 0.05 | 0.85 ± 0.06 | 0.49 | 0.56 | 4.32 ± 0.25 |
| 8439-6104 | 9.14 ± 0.16 | 1.47 ± 0.11 | 0.44 ± 0.02 | 2.86 ± 0.04 | 2.34 ± 0.04 | 0.94 ± 0.02 | 0.80 ± 0.02 | 1.04 | 1.24 | 65.99 ± 1.90 |
| 12087-12705 | 7.10 ± 0.27 | 1.15 ± 0.06 | 0.40 ± 0.04 | 2.86 ± 0.05 | 2.44 ± 0.04 | 1.02 ± 0.04 | 0.70 ± 0.04 | 0.71 | 1.34 | 27.46 ± 1.87 |

| Plate-IFU | [O II] $\lambda$3727 | [O III] $\lambda$5007 | [OI] $\lambda$6300 | H$\alpha$ | [N II] $\lambda$6584 | [S II] $\lambda$6717 | [S II] $\lambda$6731 | $\log(EW_{H\alpha})$ | c(H$\beta$) | F(H$\beta$) |
|---|---|---|---|---|---|---|---|---|---|---|
| 9029-12704 | 5.89 ± 0.21 | 1.32 ± 0.08 | 0.40 ± 0.06 | 2.86 ± 0.08 | 3.33 ± 0.08 | 1.26 ± 0.14 | 0.89 ± 0.13 | 0.56 | 0.61 | 5.50 ± 0.38 |
| 8943-12701 | 3.08 ± 0.11 | 1.21 ± 0.06 | 0.38 ± 0.05 | 2.88 ± 0.09 | 2.66 ± 0.09 | 1.08 ± 0.05 | 0.81 ± 0.05 | 0.59 | 0.30 | 2.99 ± 0.16 |
| 7978-12701 | 3.61 ± 0.14 | 0.81 ± 0.16 | 0.38 ± 0.07 | 2.91 ± 0.10 | 3.12 ± 0.08 | 1.34 ± 0.07 | 0.85 ± 0.06 | 0.79 | 0.68 | 3.36 ± 0.32 |
| 10512-6101 | 4.63 ± 0.08 | 1.63 ± 0.04 | 0.97 ± 0.07 | 2.87 ± 0.06 | 3.04 ± 0.06 | 1.40 ± 0.03 | 1.46 ± 0.03 | 0.71 | 0.42 | 13.52 ± 0.51 |
| 10842-3704 | 7.68 ± 0.13 | 1.46 ± 0.04 | 0.92 ± 0.04 | 2.89 ± 0.07 | 3.22 ± 0.06 | 1.94 ± 0.12 | 0.80 ± 0.10 | 1.27 | 0.55 | 8.53 ± 0.30 |
| 9893-6102 | 5.63 ± 0.32 | 1.28 ± 0.12 | 0.81 ± 0.13 | 2.87 ± 0.07 | 4.85 ± 0.09 | 1.36 ± 0.11 | 0.98 ± 0.11 | 0.58 | 0.70 | 2.48 ± 0.33 |
| 8146-12702 | 7.04 ± 0.20 | 1.61 ± 0.10 | 0.63 ± 0.07 | 2.88 ± 0.06 | 2.30 ± 0.07 | 0.94 ± 0.07 | 0.99 ± 0.08 | 0.53 | 0.74 | 4.08 ± 0.30 |
| 8459-3702 | 8.94 ± 0.23 | 1.27 ± 0.06 | 0.69 ± 0.07 | 2.89 ± 0.07 | 4.57 ± 0.07 | 1.22 ± 0.06 | 0.69 ± 0.06 | 0.76 | 0.91 | 12.43 ± 0.62 |
| 8978-12705 | 2.94 ± 0.07 | 1.37 ± 0.03 | 0.29 ± 0.02 | 2.87 ± 0.04 | 2.51 ± 0.05 | 0.83 ± 0.02 | 0.69 ± 0.02 | 1.01 | 0.46 | 16.72 ± 0.36 |
| 8990-9101 | 4.71 ± 0.21 | 1.69 ± 0.07 | 0.36 ± 0.04 | 2.86 ± 0.07 | 2.84 ± 0.07 | 0.97 ± 0.05 | 0.74 ± 0.05 | 0.68 | 0.74 | 5.50 ± 0.44 |